\documentclass[conference,10pt]{IEEEtran}
\IEEEoverridecommandlockouts
\usepackage{cite}
\usepackage{amsmath,amssymb,amsfonts,mathrsfs}
\usepackage{algorithmic}
\usepackage{graphicx}
\usepackage{textcomp}
\usepackage{braket}
\def\BibTeX{{\rm B\kern-.05em{\sc i\kern-.025em b}\kern-.08em
    T\kern-.1667em\lower.7ex\hbox{E}\kern-.125emX}}

\UseRawInputEncoding

\begin{document}

\title{Entropy Rate: a Route from a Gaussian Field to the Statistical Mechanics
of Thermal Radiation
\\
}

\author{Arnaldo Spalvieri \\
Dipartimento di Elettronica, Informazione e Bioingegneria \\
Politecnico di Milano\\
arnaldo.spalvieri@polimi.it \\
ORCID 0000-0002-8336-7996
}

\maketitle

\begin{abstract}

Maximum Entropy or canonical typicality, when applied to the
electromagnetic field, can provide starting points for deriving
the Bose photon-number distribution of thermal radiation, but they
rely on statistical postulates whose experimental testability is
controversial.

Here we start from the experimentally testable assumption that the
electromagnetic field is Gaussian. We analyze the field by
projecting it onto orthonormal time-frequency basis functions with
unit time-bandwidth product that represent the orthogonal
time-frequency modes of the field. Using the Glauber-Sudarshan
representation, we show that the joint density operator in the
photon-number basis factorizes into diagonal single-mode density
operators with geometric photon-number probabilities, while the
mean photon occupations of the modes remain undetermined. Applying
Clausius' relation to the total entropy and energy rates of the
independent modes, we obtain Bose's mean photon occupancy law.

Thus, our main claim is that Bose statistics are derived, for the
first time, from explicit, experimentally testable statistical
assumptions, without relying on untestable statistical postulates.

Finally, the statistical-mechanical picture is completed by
deriving the Boltzmann distribution of energy-level occupation
probabilities for the emitting and absorbing material from the
mean photon occupation numbers through microscopic detailed
balance.

\vspace{0.5cm}

{\em Keywords:}\hspace{0.2cm}{\bf Gaussian Field, Thermal
Equilibrium, Entropy, Glauber-Sudarshan Representation, Bose
Statistics, Boltzmann Statistics.}

\end{abstract}

\section{Introduction}

Entropy Maximization (MaxEnt) with expected energy constraint is
the basic assumption  from which Bose derives his famous occupancy
statistics of photons \cite{Bose1924}. Later, Jaynes
\cite{Jaynes1957} interpreted the MaxEnt principle as a general
inference principle of statistical mechanics from which the
equilibrium probability distributions are derived. The MaxEnt
principle has found widespread applications and has been supported
by axiomatic consistency arguments \cite{ShoreJohnson1980}.
However, its status as a fundamental principle of statistical
mechanics remains controversial. Friedman and Shimony
\cite{FriedmanShimony1971} and Dias and Shimony
\cite{DiasShimony1981} identified limitations of the MaxEnt
prescription.  Uffink \cite{Uffink1995}, in a comprehensive
critical review of MaxEnt and its foundations, emphasizes the
subjective Bayesian character of Jaynes' approach, in which
probabilities represent states of knowledge rather than objective
properties of physical systems. Consequently, MaxEnt, as a rule of
inference, does not by itself provide an experimentally testable
physical explanation of the statistical behavior of microscopic
systems.

Beyond the epistemological considerations discussed above, a
further issue concerns the sufficiency of the mean-energy
constraint itself. Indeed, Macovei \cite{Macovei2024} predicts
sub-Poissonian photon correlations in the radiation emitted by
small ensembles of atoms in thermal equilibrium with their
environment. Such correlations are incompatible with the geometric
photon-number statistics obtained from MaxEnt under the sole
mean-energy constraint. This suggests that, irrespective of its
epistemological justification, a physically adequate application
of MaxEnt to the emitted radiation may require additional
constraints reflecting the physical properties and dynamics of the
source, e.g. constraints on the number of radiating atoms.

An alternative statistical route to the thermal occupation-number
distribution is provided by Mag\'an and Vandoren \cite{Magan2015}.
For a massless scalar field in two dimensions, taking inspiration
from the principle of canonical typicality of \cite{Goldstein2006}
and \cite{Popescu2006} they employ Haar-random pure states in a
fixed-energy Fock subspace and obtain the reduced density matrix
of an individual mode by tracing out the remaining modes. They
show that, in the thermodynamic limit of the global field, the
occupation-number distribution of the individual mode approaches
the Bose distribution, with the temperature determined by the
energy density of the global field. However, Haar randomness is
assumed rather than derived from the microscopic dynamics of
radiation emission. In particular, the approach does not establish
that the global field states generated by a physical thermal
source are distributed according to the Haar measure. More
generally, the physical justification for choosing this measure
and the conditions under which it can emerge from microscopic
dynamics have been recognized as nontrivial issues
\cite{Dahlsten2014,Boes2018}.

The approach proposed in this paper requires neither the MaxEnt
assumption nor the Haar-uniform sampling of global pure quantum
states. Our analysis builds on the work of Glauber
\cite{Glauber1963}, who showed that, for a stationary Gaussian
field, the complex amplitudes of distinct monochromatic normal
modes are statistically independent, and derived the resulting
geometric photon-number statistics. In our framework, the Central
Limit Theorem (CLT) provides sufficient microscopic conditions on
the radiating material for the emergence of field Gaussianity.
However, Glauber's monochromatic normal modes have zero bandwidth
and no associated finite temporal duration, so his representation
does not directly provide the temporal normalization needed to
convert the sum of modal entropies into an entropy rate.

We overcome this difficulty by projecting the classical field
envelope onto an orthonormal time-frequency basis formed by
time-shifted and frequency-shifted $\mathrm{sinc}$ impulses. The
frequency shifts partition the spectrum into contiguous,
non-overlapping bands, each of arbitrarily small but nonzero
width, with a corresponding equivalent impulse duration determined
by the unit time-bandwidth product. The relationship between
photon occupation, temporal-mode duration, and photon rate has
also been established for filtered thermal radiation by Nair and
Tsang \cite{NairTsang2015}, who show that a bandwidth $\Delta\nu$
observed over a time interval $T$ supports approximately
$T\Delta\nu$ independent thermal modes, each with the same mean
photon occupation. In our construction, this correspondence is
made explicit through the equivalent duration of the
$\mathrm{sinc}$ modes before taking the continuous-frequency
limit.

 Under spectral
flatness within each frequency band, the Gaussian modal
coefficients can be treated as statistically independent
\cite{Wang2025}, making the photon-number entropies additive over
the time-frequency modes.

The sum of the modal entropies still diverges as the bandwidth
tends to zero, but, thanks to the unit time-bandwidth product, the
duration of the $\mathrm {sinc}$ basis functions diverges at the
same rate as the sum of the modal entropies, yielding a finite,
bandwidth-independent entropy rate. The entropy rate of thermal
bosonic radiation has already been studied by Pendry
\cite{Pendry1983}. Unlike Pendry \cite{Pendry1983}, who assumes
the Planckian spectrum to derive the entropy rate, we first
express the entropy rate in terms of the undetermined spectrum and
then impose Clausius' relation to determine it.

Finally, we observe that Varr\'o \cite{Varro2007} proposed a
derivation of Planck's law from Gaussian field statistics without
invoking any axiomatic assumption. Starting from the exponential
energy distribution of a classical Gaussian field, Varr\'o obtains
the geometric distribution by taking the integer part of the
classical energy expressed in units of $h\nu$. However, Varr\'o's
identification of photon number with the integer part of the
classical modal energy differs fundamentally from the Glauber
approach adopted here. In Varr\'o's construction, the
discretization leaves a random fractional energy remainder,
interpreted as an unobserved ``dark'' energy. In the
Glauber-Sudarshan representation, instead, the classical modal
energy determines the mean of the conditional Poisson
photon-number distribution associated with a coherent state. The
resulting statistical mixture of coherent states yields the
geometric distribution while preserving the equality between the
mean classical field energy and the mean photon energy, without
introducing any residual energy contribution.


\section{The Gaussian field and
its entropy}

In the following, we consider an electromagnetic field propagating
in a single direction and polarization, within a single transverse
spatial mode. Let $\Sigma$ denote the full transverse domain in a
plane perpendicular to the propagation direction. The transverse
field profile is normalized with respect to the power flux
integrated over $\Sigma$, so that the corresponding real classical
field amplitude $x(t)$ has squared amplitude $x^2(t)$ equal to the
total instantaneous electromagnetic power crossing $\Sigma$. We
denote by $\Psi(\nu)$ the initially unknown one-sided energy
spectral density of this radiation, defined as the mean
electromagnetic energy per unit propagation length and per unit
frequency, integrated over the same transverse domain $\Sigma$.
The material presented in this section is based on the classical
and experimentally testable assumption that
\begin{itemize} \item $x(t)$ is a zero-mean
stationary Gaussian random process with circular Gaussian complex
envelopes and with smooth $\Psi(\nu)$.
\end{itemize}

For $j=\cdots -1,0,1 \cdots,$ let us divide the spectrum into the
frequency slots $\nu_j-\Delta \nu/2 \leq \nu < \nu_j +\Delta
\nu/2$ with $\nu_j=j \Delta \nu + \Delta \nu/2$. Let $\tilde
x_j(t)$ be the complex envelope of $x(t)$ around frequency $\nu_j$
and, for $k=\cdots -1,0,1 \cdots$ let us represent $x(t)$ by the
following samples of the complex envelopes taken at $t_k=k \Delta
\nu^{-1}$:
\begin{equation}\tilde{x}_{j,k}=\int_{-\infty}^{\infty} x(\tau)
\phi^*_{j,k}(\tau)d\tau,\label{an} \end{equation} where the
orthonormal basis functions that encode the time-frequency
orthogonal modes of $x(t)$ are
\begin{equation}\phi_{j,k}(\tau)=\sqrt{\Delta \nu} \ \mathrm{sinc}(k
-\Delta \nu\tau) \exp(i 2 \pi \nu_j \tau), \nonumber 
\end{equation} with
\[\mathrm{sinc}(\tau)=\frac{\sin(\pi \tau)}{\pi \tau}.\]
The equivalent duration of each $\mathrm{sinc}$ basis function is
$1/\Delta \nu$ and its equivalent bandwidth is $\Delta \nu$, so
the time-bandwidth product is unity, regardless of the value of
$\Delta \nu$. The reconstruction formula is
\begin{equation}x(t)=\sum_j \sum_k \tilde x_{j,k} \phi_{j,k}(t).\label{syn}\end{equation}


Let $\Delta \nu$ be sufficiently small that $\Psi(\nu)$ can be
regarded as constant within each frequency slot. Since the Fourier
transform of the $ \mathrm{sinc}$ function is also constant inside
the same bandwidth and zero outside, the samples $\tilde x_{j,k}$
for $j \geq 0$ are uncorrelated in time and frequency and, by
Gaussianity of the field, they are also mutually independent.

In principle, the above construction can be extended to complete
orthonormal time-frequency representations other than the
$\mathrm{sinc}$-based one formulated in (\ref{an})-(\ref{syn}),
provided that the statistical independence of the corresponding
Gaussian coefficients is preserved. Orthogonal wave-packet
decompositions of electromagnetic radiation into independent
time-frequency modes under spectral-flatness conditions have been
considered in \cite{Wang2025} in the context of thermal radiation.
Note that only local spectral flatness is required for statistical
independence, while no specific thermal form of the global
spectrum needs to be assumed, since this form will be determined
by imposing Clausius' relation.

Owing to the stationarity of the field, all time samples within a
given frequency slot are statistically equivalent. Therefore, in
the following analysis, we omit the discrete time subscript $k$
for brevity and characterize the generic sample $\tilde x_j$ as a
zero-mean complex circular Gaussian random variable, whose
variance depends only on $\nu_j$. Since the basis functions have
unit squared norm, the squared modulus of each sample represents
its contribution to the field energy. Therefore, its expectation,
equal to the variance of the circular complex Gaussian random
variable $\tilde x_j$, is, for propagation in the vacuum,
\[\mathbf E\{|\tilde x_j|^2\}=\mathbf E\{|\tilde
x_{-j-1}|^2\}=\frac{c\Psi(\nu_j)}{2}, \ \ j=0,1, \cdots,\] where
$c$ is the speed of the light and the division by two arises
because $\Psi(\nu)$ is one-sided.


We now invoke the well-established Glauber-Sudarshan
\cite{Glauber1963,Sudarshan1963} representation of the quantized
electromagnetic field to describe the photon statistics associated
with the independent time-frequency modes introduced above.
Specifically, we represent the quantum field in the $j$-th
frequency slot in the positive-frequency range by a statistical
mixture of coherent states $\ket{\alpha_j}$, which are eigenstates
of the corresponding annihilation operator, with complex
amplitudes $\alpha_j$ distributed according to the Gaussian
statistics of the classical field. In the following, we express
both energy and temperature in kelvin \cite{Olmsted2010}, using
Boltzmann's constant to convert energy from joules to kelvin. With
this convention, one kelvin corresponds to $1.38 \cdot 10^{-23}$
joule. Accordingly, we introduce the single quantum-thermal
constant
\[
\eta=\frac{h}{k_{\rm B}}=4.7992 \cdot 10^{-11}, \qquad
[\eta]=\mathrm{K\,s},
\]
where $h$ is Planck's constant and $k_{\rm B}$ is Boltzmann's
constant. With this convention, the mean photon energy in the
$j$-th frequency slot is $\eta\nu_j$. By equating the classical
modal energy to the mean energy of the corresponding coherent
state and identifying the phase of $\alpha_j$ with that of $\tilde
x_j$, we obtain the dimensionless complex amplitude
\begin{equation}
\alpha_{j}= \frac{\tilde x_j + \tilde x^*_{-j-1}}{\sqrt{2 \eta
\nu_j}}=\tilde x_{j} \sqrt{\frac{2}{\eta\nu_j}} , \ \ j=0,1,
\cdots. \nonumber 
\end{equation}

 The density operator that represents the mixed state of each mode is
diagonal in the photon-number basis, with geometric diagonal
elements \cite{Glauber1963}:
\begin{equation}
\hat\rho_j = \sum_{n=0}^{\infty} \frac{\bar N_j^n}{(1+\bar
N_j)^{n+1}} |n\rangle\langle n|, \label{rho}
\end{equation}
where, from the definition of $\alpha_j$, the mean photon number
is
\begin{equation}
\bar N_j=\mathbf E\{N_j\} =\frac{c\Psi(\nu_j)}{\eta\nu_j}.
\label{celloccupation} \end{equation} The von Neumann entropy of
the diagonal density operator $\rho_j$ coincides with the Shannon
entropy of the geometric distribution:
\begin{align}
H_j &=-\operatorname{Tr}
(\hat\rho_j\ln\hat\rho_j)\nonumber\\
&=\left[ (1+\bar N_j)\ln(1+\bar N_j) -\bar N_j\ln\bar N_j \right],
\label{cellentropy}
\end{align}
where natural logarithms are used, and, since in this paper
entropy is dimensionless, we adopt the symbol $H$ commonly used in
information theory.


\section{Bose occupancy law and the canonical density operator of the field}

Consider a radiation field thermalized by a source at temperature
$T$. We assume that the source has nonzero spectral emissivity at
every frequency under consideration, allowing all field modes to
exchange energy with the source and reach thermal equilibrium. The
initially unknown spectral density $\Psi(\nu)$ can be determined
by imposing Clausius' relation to the the total entropy and energy
rates of field's modes.

As discussed in the previous section, the classical Gaussian
amplitudes of distinct modes are statistically independent.
Therefore, the joint density operator associated with the field
modes factorizes into a product of single-mode density operators.
As a consequence, the total entropy $H$ and the total mean energy
$E$ are additive over the time-frequency modes. For a fixed time
index, summing over all frequency slots gives:
\begin{equation}H=\sum_{j=0}^{\infty}H_j, \ \ E=\sum_{j=0}^{\infty}
c \Psi(\nu_j). \nonumber 
\end{equation}

 However, the above two sums
depend on the arbitrary $\Delta \nu$. Specifically, they are
proportional to $1/\Delta \nu$ because the equivalent time
duration of the basis functions is proportional to $1/\Delta \nu$.
The quantities that are independent of the arbitrary bandwidth
$\Delta \nu$ are the entropy rate $\dot H$ and the mean power
(energy rate) $\dot E$ representing entropy and mean energy
transported per unit time through the transverse cross-section. We
define them by taking the limit for $\Delta \nu \rightarrow 0$,
thus allowing for arbitrarily fine frequency resolution:
\begin{align}
\dot H &=\lim_{\Delta\nu\to0} \Delta\nu\sum_{j=0}^{\infty}H_j
=\int_0^\infty H(\nu)\,d\nu,
\label{hrate}\\
\dot E &=\lim_{\Delta\nu\to0} \Delta\nu\sum_{j=0}^{\infty}c
\Psi(\nu_j) =\int_0^\infty c \Psi(\nu)\,d\nu. \label{erate}
\end{align}
Since $H(\nu)$ and $c \Psi(\nu)$ are associated to basis functions
having a unit time-bandwidth product, they can be interpreted as
entropy rate spectral density and energy rate (power) spectral
density, respectively.

The corresponding mean photon occupation $\bar
N(\nu)=c\Psi(\nu)/(\eta\nu)$ is obtained as the
continuum-frequency limit of the previously defined $\bar N_j$.
This limit concerns the statistical properties of the
finite-bandwidth modes, rather than the existence of a mode of
strictly zero bandwidth. For each finite $\Delta\nu$, a mode with
mean photon occupation $\bar N_j$ has an equivalent duration
$1/\Delta\nu$ and contributes $\bar N_j\Delta\nu$ to the mean
photon rate. Consequently, $\bar N(\nu)d\nu$ represents the
contribution of the infinitesimal frequency interval $[\nu-d
\nu/2,\nu+d\nu/2)$ to the mean photon rate.

Both Glauber \cite{Glauber1963} and Pendry \cite{Pendry1983}
describe radiation in terms of monochromatic modes with purely
sinusoidal time dependence, without explicitly associating them
with a finite temporal duration. For each monochromatic mode,
Glauber identifies the mean photon occupation with the expectation
of the squared modulus of its time-independent complex amplitude
and derives the geometric photon-number distribution from the
Gaussian statistics of that amplitude. Pendry, instead, takes the
mean photon occupation as given and relates it to the photon flux
through the propagation velocity. In contrast, consistent with the
time-frequency interpretation discussed by Nair and Tsang
\cite{NairTsang2015}, our finite-bandwidth decomposition
explicitly connects mean photon occupation, equivalent mode
duration, and photon rate before taking the continuous-frequency
limit.

We now impose Clausius' relation at thermal equilibrium for
arbitrary independent variations of the mean photon occupation
numbers at different frequencies. This requires, at each
frequency,
\begin{equation}\left. \frac{d
\dot H}{d \dot E}\right|_{\nu}=\frac{1}{c} \frac{d H(\nu)/d \bar
N(\nu)}{d \Psi(\nu)/d \bar N(\nu)} =\frac{1}{T}. \nonumber
\end{equation}
Using (\ref{celloccupation}) and (\ref{cellentropy}) we obtain
\begin{equation}
\bar N(\nu)=\frac{1}{\exp(\eta\nu/T)-1} \label{nx},
\end{equation}
that is the Bose occupancy law. The entropy rate spectral density
and the power spectral density are, respectively,
\begin{align}
&H(\nu)=\frac{\eta \nu}{T (\exp(\eta \nu/T)-1)} -\ln(1-\exp(-\eta
\nu/T)),\nonumber \\ & c\Psi(\nu) =
\frac{\eta\nu}{\exp(\eta\nu/T)-1}.
\nonumber 
\end{align}
The last equation gives the quantum Nyquist expression for the
available thermal power spectral density in a single propagating
channel \cite{Nyquist1928}, while $\Psi(\nu)$ is Planck's energy
spectral density per unit length for the one-dimensional channel
considered here. The extension to three spatial degrees of freedom
is standard. Substituting the preceding expressions into
(\ref{hrate}) and (\ref{erate}) yields
\begin{align}
& \dot H=\frac{\pi^2 T}{3 \eta}, \ \  \dot E= \frac{\pi^2 T^2}{6
\eta},
\nonumber 
\end{align}
yielding
\[\dot H = \pi \sqrt{\frac{2 \dot E}{3 \eta}},\]
which, under our convention of expressing energy in kelvin, is
consistent with eqn. (29) of \cite{Pendry1983}.

Substituting the mean occupation number (\ref{nx}) into the
geometric density operator (\ref{rho}), yields the canonical Gibbs
density operator of a single bosonic mode:
\[ \hat{\rho}(\nu)=(1-e^{-\eta\nu/T})e^{-\hat H(\nu)/T}, \] where
$\hat H(\nu)=\eta\nu\sum_{n=0}^{\infty}n\ket{n}\bra{n}$ is the
single-mode Hamiltonian operator, with the zero-point energy
omitted. For independent modes, the total field state therefore
takes the Gibbs form associated with the free-field Hamiltonian.

\section{Boltzmann distribution}

We now complete the statistical-mechanical description of thermal
radiation by deriving the Boltzmann distribution of energy-level
occupation probabilities for the emitting and absorbing material
from the mean photon occupation numbers obtained in the preceding
sections. In the classical Einstein \cite{Einstein1917} and
Feynman \cite{FeynmanIII} treatments, the Boltzmann distribution
of the material is assumed, then, from it, the equilibrium
radiation spectrum is derived through detailed balance between
absorption, stimulated emission, and spontaneous emission. Here,
the logical path is reversed. Moreover, rather than considering
transitions between two energy levels coupled to a single
radiation frequency, we describe the interaction between the
continuous radiation spectrum and the spectral absorptivity of the
material, allowing for the superposition and broadening of its
absorption and emission lines. Microscopic detailed balance then
determines the equilibrium populations of the material from the
independently derived photon occupation statistics.

We adopt the standard model of photon absorption and emission
associated with energy transitions occurring independently in
different particles of the material. The photon frequency
$\nu_{h,l}$, corresponding to a transition between the higher
energy level $\epsilon_h$ and the lower energy level $\epsilon_l$,
satisfies
\[\eta \nu_{h,l}= \epsilon_{h}-\epsilon_{l}.\]
Absorption and emission are not instantaneous; rather they are
associated with the same narrowband spectral line, sharply peaked
around $\nu_{h,l}$ \cite{FeynmanLectures1}.

According to Einstein's theory of radiative transitions
\cite{Einstein1917}, the mean rates of absorption and stimulated
emission are proportional to the mean intensity of the incident
radiation and hence, in our continuum-frequency representation, to
$\bar N(\nu)$, whereas spontaneous emission occurs even in the
absence of incident radiation. Feynman's quantum-mechanical
treatment \cite{FeynmanIII} shows that, for a radiation state
containing $n$ photons, the probabilities of absorption and
emission at equilibrium are proportional to $n$ and $n+1$,
respectively, where the additional term of unity represents
spontaneous emission and follows from the bosonic character of
photons. Assuming that, at thermal equilibrium, $\bar N(\nu)$
given by (\ref{nx}) is approximately constant across the
narrowband spectral line associated to the energy transition, with
value $\bar N(\nu_{h,l})$, the mean numbers of upward and downward
transitions are proportional to
\[\bar N_{l \rightarrow h}(\nu_{h,l}) \propto \bar M_{l}\bar N(\nu_{h,l})
\int_0^\infty A_{h,l}(\nu)d\nu \] and \[ \bar N_{h \rightarrow
l}(\nu_{h,l}) \propto \bar M_{h}(1+\bar N(\nu_{h,l}))\int_0^\infty
R_{h,l}(\nu)d\nu,
\] respectively, where $\bar M_{l}$ and $\bar M_{h}$ denote
the mean numbers of particles occupying the lower and higher
energy levels of the nondegenerate transition, while
$A_{h,l}(\nu)$ and $R_{h,l}(\nu)$ denote the corresponding
spectral contributions of the energy transition to the total
absorptivity and radiativity, respectively. By Kirchhoff's law,
radiativity and absorptivity are equal at thermal equilibrium.
Consequently, the same spectral integral appears in both
expressions and cancels when the upward and downward transition
rates are equated:
\begin{equation}\frac{\bar M_{h}}{\bar M_{l}}= \frac {\bar N(\nu_{h,l})}{1+\bar N(\nu_{h,l})}
= \exp(-\eta \nu_{h,l} /T) . \nonumber 
\end{equation} Importantly, this derivation does not require
$A_{h,l}(\nu)$ to be constant over the narrowband spectral line
and remains valid even when the absorbing material allows for only
one energy transition. Since the mean atomic populations are
proportional to the corresponding single-atom occupation
probabilities, the equilibrium balance of the mean absorption and
emission rates yields the Boltzmann ratio:
\[ \frac{P(\epsilon_{h})}{P(\epsilon_{l})} = \exp(-\eta \nu_{h,l} /T)=\exp(-(\epsilon_h-\epsilon_l)/T).\]

Assuming that the allowed radiative transitions connect all the
energy levels $\{\epsilon_k\}$ of the material and that detailed
balance holds for each allowed transition, the above relation
implies that $P(\epsilon_{k})\exp(\epsilon_k/T)$ is independent of
$k$. Normalization then yields the Boltzmann distribution in its
standard form:
\[P(\epsilon_k)=\frac{\exp(-\epsilon_k/T)}{\sum_{j}\exp(-\epsilon_j/T)}.\]

\section{Conclusion}

There has long been a debate over the maximum-entropy principle
under a mean-energy constraint and the assumption of a uniform
measure over the energy-constrained set of global states in
statistical mechanics. These assumptions remain controversial
because it is still unclear how they can be tested experimentally.
This paper derives, for the first time, two fundamental
distributions of statistical mechanics, namely the Bose
photon-number distribution and the Boltzmann distribution for
massive particles at thermal equilibrium, starting from the
experimentally testable assumption of Gaussianity of the
electromagnetic field.

Remarkably, the circular Gaussian and the geometric distributions
of the preceding sections are maximum-entropy distributions under
their respective mean-energy constraints. However, in the present
approach, these distributions are not selected by the MaxEnt
principle under a mean-energy constraint. In particular, when the
conditions of the CLT are not satisfied, statistically independent
geometric photon-number distributions are no longer guaranteed,
even when the emitting atoms are in thermal equilibrium with their
environment. This limitation is illustrated by Macovei
\cite{Macovei2024}, who predicts non-geometric photon statistics
for radiation spontaneously emitted by small ensembles of atoms in
thermal equilibrium with a thermal bath. Such statistics cannot be
reproduced by applying MaxEnt to the radiation field under the
sole constraint of its mean energy.



\begin{thebibliography}{999}

\bibitem{Bose1924}
S. N. Bose, "Planck's Law and the Light Quantum Hypothesis,"
translated by O. Theimer and B. Ram, in "The Beginning of Quantum
Statistics," American Journal of Physics, vol. 44, no. 11, pp.
1056--1057, 1976. Original work published in 1924.
doi:10.1119/1.10584.

\bibitem{Jaynes1957}
E. T. Jaynes, "Information Theory and Statistical Mechanics,"
Physical Review, vol. 106, pp. 620--630, 1957.
doi:10.1103/PhysRev.106.620.

\bibitem{ShoreJohnson1980}
J. E. Shore and R. W. Johnson, "Axiomatic Derivation of the
Principle of Maximum Entropy and the Principle of Minimum
Cross-Entropy," IEEE Transactions on Information Theory, vol. 26,
no. 1, pp. 26--37, 1980. doi:10.1109/TIT.1980.1056144.

\bibitem{FriedmanShimony1971}
K. Friedman and A. Shimony, "Jaynes's Maximum Entropy Prescription
and Probability Theory," Journal of Statistical Physics, vol. 3,
pp. 381--384, 1971. doi:10.1007/BF01008275.

\bibitem{DiasShimony1981}
P. M. C. Dias and A. Shimony, "A Critique of Jaynes' Maximum
Entropy Principle," Advances in Applied Mathematics, vol. 2, no.
2, pp. 172--211, 1981. doi:10.1016/0196-8858(81)90003-8.

\bibitem{Uffink1995}
J. Uffink, "Can the Maximum Entropy Principle Be Explained as a
Consistency Requirement?" Studies in History and Philosophy of
Modern Physics, vol. 26, no. 3, pp. 223--261, 1995.
doi:10.1016/1355-2198(95)00015-1.

\bibitem{Macovei2024}
M. A. Macovei, "Dense dipole-dipole-coupled two-level systems in a
thermal bath," Physical Review A {\bf 110}, 023712 (2024).
doi:10.1103/PhysRevA.110.023712.



\bibitem{Magan2015}
J. M. Mag\'an and S. Vandoren, "Entanglement in Fock space of
random QFT states," Journal of High Energy Physics {\bf 2015}, 150
(2015). doi:10.1007/JHEP07(2015)150.

\bibitem{Goldstein2006}
S. Goldstein, J. L. Lebowitz, R. Tumulka, and N. Zangh\`i,
"Canonical Typicality," Phys. Rev. Lett., vol. 96, no. 5, Art. no.
050403, 2006. doi:10.1103/PhysRevLett.96.050403.

\bibitem{Popescu2006}
S. Popescu, A. J. Short, and A. Winter, "Entanglement and the
Foundations of Statistical Mechanics," Nature Physics, vol. 2, no.
11, pp. 754--758, 2006. doi:10.1038/nphys444.



\bibitem{Dahlsten2014}
O. C. O. Dahlsten, C. Lupo, S. Mancini, and A. Serafini,
Entanglement typicality, J. Phys. A: Math. Theor. \textbf{47},
363001 (2014). https://doi.org/10.1088/1751-8113/47/36/363001

\bibitem{Boes2018}
P. Boes, H. Wilming, J. Eisert, and R. Gallego, "Statistical
Ensembles without Typicality," Nature Communications, vol. 9,
article 1022, 2018. doi:10.1038/s41467-018-03230-y.




\bibitem{Glauber1963}
R. J. Glauber, "Coherent and Incoherent States of the Radiation
Field," Physical Review, vol. 131, pp. 2766--2788 (1963).
https://doi.org/10.1103/PhysRev.131.2766







\bibitem{NairTsang2015}
R. Nair and M. Tsang, ``Quantum optimality of photon counting for
temperature measurement of thermal astronomical sources,''
Astrophys. J. \textbf{808}, 125 (2015).
https://doi.org/10.1088/0004-637X/808/2/125



\bibitem{Wang2025}
Y. Wang, Y. Zhang, and V. O. Lorenz, "Temporally localized quantum
operations on continuous-wave thermal light," Physical Review
Letters {\bf 135}, 113602 (2025). doi:10.1103/y4v8-1wgm.

\bibitem{Pendry1983}
J. B. Pendry, Quantum limits to the flow of information and
entropy, J. Phys. A: Math. Gen. \textbf{16}, 2161--2171 (1983).
https://doi.org/10.1088/0305-4470/16/10/012

\bibitem{Varro2007}
S. Varr\'o, ``Irreducible decomposition of Gaussian distributions
and the spectrum of black-body radiation,'' Physica Scripta
\textbf{75}, 160--169 (2007).
https://doi.org/10.1088/0031-8949/75/2/008



\bibitem{Sudarshan1963}
E. C. G. Sudarshan, "Equivalence of Semiclassical and Quantum
Mechanical Descriptions of Statistical Light Beams," Physical
Review Letters, vol. 10, pp. 277--279 (1963).
https://doi.org/10.1103/PhysRevLett.10.277

%
\bibitem{Olmsted2010}
J.~Olmsted III, ``Rescaling temperature and entropy,''
\textit{Journal of Chemical Education}, vol.~87, pp.~1195--1199,
2010.




%
%








%





\bibitem{Nyquist1928}
H. Nyquist, "Thermal agitation of electric charge in conductors,"
Physical Review {\bf 32}, 110--113 (1928).
doi:10.1103/PhysRev.32.110.





\bibitem{Einstein1917}
A. Einstein, "On the Quantum Theory of Radiation," in The Old
Quantum Theory, D. ter Haar, Ed., Pergamon Press, Oxford, 1967,
pp. 167--183. English translation of the original paper published
in Physikalische Zeitschrift, vol. 18, pp. 121--128, 1917.

\bibitem{FeynmanIII}
R. P. Feynman, R. B. Leighton, and M. Sands, \textit{The Feynman
Lectures on Physics}, Vol. III, Chap. 4, Secs. 4-4 and 4-5,
Addison-Wesley, Reading, MA (1965).


\bibitem{FeynmanLectures1}
R. P. Feynman, R. B. Leighton, and M. Sands, \textit{The Feynman
Lectures on Physics}, Vol. I, Chap. 41, Secs. 41-2 and 41-3,
Addison-Wesley, Reading, MA, 1963.




%


















%


















%
%
%
%
%
%
%
%
%
%
%
%
%
%
%
%
%
%
%
%
%
%
%
%
%
%
%
%
%
%
%
%
%
%
%
%
%
%
%
%
%
%
%
%
%
%
%
%
%
%
%
%
%
%
%
%
%
%
%
%
%
%
%
%
%
%
%
%
%
%
%
%
%
%
%
%
%
%
%
%
%
%
%
%
%
%
%
%
%
%
%
%
%
%
%
%
%
%
%
%
%
%
%
%
%
%
%
%
%
%
%
%
%
%
%
%
%
%
%
%
%
%
%
%
%
%
%
%
%
%
%
%
%
%
%
%
%
%
%
%
%
%
%
%
%
%
%
%
%
%
%
%
%
%
%
%
%
%
%
%
%
%
%
%
%
%
%

\end{thebibliography}
\end{document}